\documentclass[a4paper]{jpconf}
\usepackage{amssymb}
\usepackage{amsmath}
\usepackage{amsthm}
\usepackage{enumerate}
\newcommand{\be}{\begin{equation}}

\newcommand{\ee}{\end{equation}}
\newcommand{\bea}{\begin{eqnarray}}
\newcommand{\eea}{\end{eqnarray}}
\newcommand{\coleps}{\varepsilon^{\alpha\beta I}}
\newcommand{\flaeps}{\epsilon_{ij I}}
\newcommand{\apj}{{\rm ApJ}}	
\newtheorem{defs}{Definition}[section]
\usepackage{graphicx}
\usepackage{bm}

\begin{document}
\title{The amazing properties of crystalline color superconductors}

\author{Massimo Mannarelli}

\address{INFN, Laboratori Nazionali del Gran Sasso, Assergi (AQ), Italy}

\ead{massimo@lngs.infn.it}

\begin{abstract}
This paper is a brief journey into the amazing realm of crystalline color superconductors. 
Starting from a qualitative description of superfluids,  superconductors and supersolids, we show how inhomogeneous phases may arise when the system is under stress. These basic concepts are then extended to quark matter, in which a richer variety of phases can be realized. Then, the most interesting properties of the crystalline color superconductors are presented. This brief journey ends with a  discussion  of crystalline color superconductors in compact stars and related   astrophysical observables. We aim at providing a pedagogical introduction for nonexpert in the field to a few interesting properties of crystalline color superconductors, without discussing the methods and the technicalities.  Thus, the  results are presented without a proof. However, we try to give a qualitatively clear description of the main concepts, using standard  quantum field theory and analogies with condensed matter systems.  

\end{abstract}

\section{Introduction}

The crystalline color superconducting (CCSC) phase can be qualitatively described as a system of deconfined quarks in which the strong interaction favors the formation of a quark-quark condensate that is spatially modulated as a crystal. 

This rather odd phase of matter is one of the candidate phases of beta equilibrated and electrically neutral quark matter at very large baryonic density and sufficiently low temperature, see~\cite{Anglani:2013gfu} for a review.   
The appropriate conditions may be realized within  compact stellar objects (CSOs), which are  stars having a radius of about $10$ km and a  mass  of $1$ - $2$ M$_\odot$.  Soon after their birth from the gravitational collapse of the remnants of a supernova explosion, the temperature of CSOs is rather hot, of the order of $10^{11}$K, but then they rapidly cool down by neutrino emission to temperatures of the order of   $10^7$K - $10^8$K that is of the order of tens of keV at most. Although very high for our daily lives, this is  is a rather cold temperature as compared to the natural scale of Quantum Chromodynamics (QCD), $\Lambda_{\text QCD} \sim 200$  MeV. 
The density of the CSOs is also very large, it is so large that the average distance between nuclei is  less than  $1$ fm $\sim 1/\Lambda_{\it QCD}$, which is the typical size of a nucleon. Although first principle calculations are not feasible at this scale,  it is clear that for matter  squeezed at such extremes, talking about nucleons is not appropriate. In these conditions the relevant degrees of freedom should be quarks and gluons.

\begin{figure}[h!]
\begin{center}
\includegraphics[width=12cm]{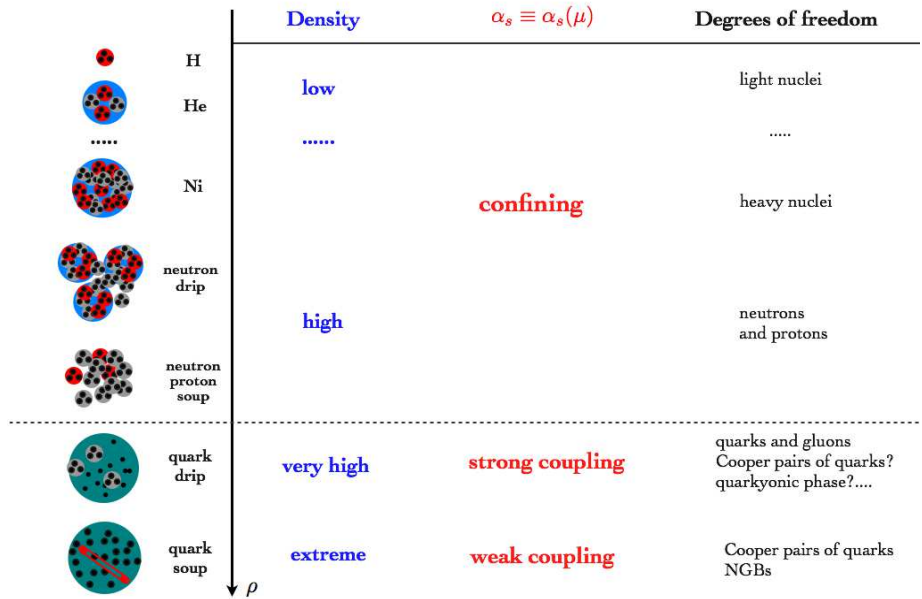}
\end{center}
\caption{\label{fig:density} Schematic representation of the  phases of baryonic matter at various densities. }
\end{figure}

A pictorial description of the behavior of matter with increasing baryonic density is given in Fig.~\ref{fig:density}. The density increases from  the top to bottom of the figure and the corresponding qualitative features of matter are sketched. In the confined phase (above the horizontal dashed line), with increasing  matter density  the number of nucleons in a  nucleus increases, reaching the so-called nuclear saturation density for heavy nuclei. Further increasing the density one expects that  nuclei start to melt and neutrons are liberated, neutron drip. Up to this point QCD is a confining theory, meaning that quarks and gluons are not the correct degrees of freedom for the description of matter. For larger values of the density the strong interaction is no more able to confine quarks,  and the ``quark drip" should happen. Note that at this point  the strong interaction is still nonperturbative. This is one of the very important aspects: there is a range of density in which quark matter if liberated is still  nonperturbatively interacting. It is  this range of densities that is expected to be relevant for compact stars. 

Only at asymptotic densities  QCD becomes perturbative and solid results based on perturbation theory can be achieved. Indeed, it is only at extreme densities that we know the state of quark  matter: it becomes a color superconductor. In this phase quarks form a degenerate soup, filing the  Fermi sphere up to a large Fermi energy. The color attractive interaction induces the formation of Cooper paris, with an average correlation length larger then the average distance between two quarks (bottom line of Fig.~\ref{fig:density}). 

The hope is that from the asymptotic density down to CSOs densities no other phase pops up. This is one of the uncontrollable hypothesis about color superconductors. In QCD, nonperturbative physics is typically studied by
lattice QCD simulations. Unfortunately these simulations cannot be easily done with a large baryonic chemical potential, because of the so-called sign problem~\cite{Barbour:1986jf}, see for example~\cite{Aarts:2013naa} for a recent review on progress in the field. 

In the following we shall  simply assume that  reducing the density from the asymptotic value down to  the nuclear matter phase transition, the color superconducting phase is  the only energetically favored phase. 

Before turning to a more detailed description of color superconductors, let us first define what are  superfluids and superconductors.

\section{Superfluids, superconductors and supersolids}
We  give two different definitions of  superfluids and superconductors: one  operative  and one formal.
 \begin{defs}\label{def:superfluid}
Superfluid. \underline{Operative}: Frictionless fluid with potential flow that  when rotated is threaded by quantized vortices. \underline{Formal}: System with the spontaneous breaking of a global symmetry below a critical temperature. 
\end{defs}

\noindent 
\begin{defs}\label{def:superconductor}
Superconductor. \underline{Operative}: Almost perfect diamagnet, screening  the magnetic field in restricted domains.
 \underline{Formal}:  System with the ``spontaneous breaking" of a gauge symmetry below a critical temperature.
\end{defs}

In both cases, the two definitions are equivalent. In superfluids the theoretical definition is based on the  Goldstone theorem implying that there is at least a  massless mode, $\phi$, allowing the easy transport of the quantum numbers associated with the broken global symmetries. The fluid velocity $\bm v$ can be written as $\bm v = \nabla \phi$. It follows that the vorticity,  $\nabla \times \bm v$, vanishes almost everywhere. It can be shown that the  circulation of $\bm v$ is quantized, meaning that each vortex carries a quantum of angular momentum. In superconductors, the magnetic field is screened over a distance $1/M$, with $M$ the gauge fields magnetic mass acquired by the Higgs-Anderson mechanism.  

In the analysis of the properties of a system one must verify that all the above requirements are met. As a counterexample, consider the ballistic propagation of a dilute system of neutral particles, say neutrons or neutrinos. One can describe it as a frictionless fluid with vanishing vorticity. However,  it is certainly not a superfluid. Indeed, if one could spin it, for example by confining neutrons in a  trap and then putting the trap in rotation, one would not observe quantized vortices. Moreover, no condensation occurs and no global symmetry of the system is spontaneously broken. 

Consider instead the case of neutrons forming a $nn$ condensate. This is likely to  happen in the inner crust of neutron stars. In this case the global symmetry corresponding to neutron number conservation is broken (the neutron number is not conserved because neutrons can ``disappear'' in the condensate). This system is  a  genuine superfluid. That is, the spinning neutron superfluid is believed to be threaded by quantized vortices.

There is a number of subtleties that we have not addressed in the definitions of superfluids and superconductors, however, our definitions should be sufficiently general to encompass most of the known forms of superfluids and superconductors and, more important, to avoid confusion. 

Finally, we say that a system has a superflow mode whenever there is a breaking of a symmetry of the system which allows the easy transport of a quantum number. We now wonder whether superfluids and superconductors are the only forms of superflows that are realizable in nature. 

\subsection{Supersolids}
It might sound unreasonable that a solid could become superfluid or that a superfluid could solidify keeping zero viscosity. The two concepts: superfluidity and rigidness seem to be in conflict. A solid is difficult to deform (it has by definition a nonvanishing shear modulus), a fluid has no shear modulus and a superfluid has also a vanishing shear viscosity!  Moreover,  a superfluid is a highly delocalized system while a solid is characterized by long-range order~\cite{1956PhRv..104..576P}.

Transforming a superfluid in a solid is  in general possible. A notable example is helium. Both He$^4$ (bosonic) and    He$^3$ (fermionic) become solids with increasing pressure.   Thus, the way in which helium becomes solidifies is  by compression. When compressed, the repulsion between helium atoms forces the system to a closely packed configuration, typically an hexagonal close-packed (HCP) crystal, a body centered cubic (BCC) crystal or a face-centered cubic (FCC) crystal. It is clear that once one of  these configurations is reached, helium atoms cannot easily move around. Indeed the system loses its superfluid property. However, if  defects  are present, atoms  might be able to move. For this reason it has been supposed that ``nonperfect" solid helium could be a supersolid.  However,   experimental results with helium atoms have not unambiguously shown that supersolid helium exists, see \cite{2012RvMP...84..759B, 2011arXiv1110.1323D} for reviews. Different systems consisting of ultracold bosonic  atoms with a particular long-range behavior~\cite{Henkel} seem to be good candidates~\cite{Pupillo, Pupillo2}. 

Here is a definition of supersolids (see \cite{2012RvMP...84..759B}  for more details):
\begin{defs}\label{def:supersolids}
  A supersolid is a system in which the spontaneous breaking of rotational symmetries down to a discrete symmetry and the spontaneous  breaking  of an internal symmetry   take place simultaneously and for the same type of particles.
 \end{defs} 
In a supersolid  particles are delocalized throughout the whole system, and simultaneously  there is long-range order~\cite{2012RvMP...84..759B} (meaning that the diffraction pattern of a supersolid has narrow peaks). The former property is a consequence of the breaking of an internal symmetry of the system;  the latter refers to the breaking of space symmetries. What is crucial in the above definition  is that the two spontaneous symmetry breaking  must appear simultaneously  and for the same type  of particles~\cite{2012RvMP...84..759B}.

\section{Color Superconductors}
We now turn to quark matter, see \cite{Alford:2007xm,Anglani:2013gfu} for reviews. Only up, down and strange quarks are relevant, because other quark flavors have masses larger that the typical chemical potential realizable in CSOs, of the order of $400$ MeV. The formation of quark Cooper pairs can be described by the condensate 
\begin{equation}
\langle 0|\psi_{iL}^\alpha\psi_{jL}^\beta|0\rangle=-\langle
0|\psi_{iR}^\alpha\psi_{jR}^\beta|0\rangle \propto
\sum_{I=1}^3 \Delta_{I} \coleps\flaeps\,,
\label{eq:condensate-CFL}\end{equation}
where $\psi_{iL/R}^\alpha$ represents a left/right handed quark fields with  color $\alpha$ and flavor $i$. The quantities
$\Delta_{1}$, $\Delta_{2}$ and $\Delta_{3}$, describe $d$-$s$, $u$-$s$ and $u$-$d$ Cooper pairing, respectively, and are proportional to the energy needed to break the corresponding Cooper pair;  $\varepsilon^{\alpha\beta\gamma}$ and $\epsilon_{ijk}$ are the completely antisymmetric symbols in color and flavor space, respectively. 
The color structure is determined by the fact that the completely antisymmetric $\bar 3$ channel is attractive, then requiring that the condensate has zero total angular momentum the flavor structure must be completely antisymmetric as well. For the sake of notation we have suppressed spinorial  indices and for the sake of simplicity we have neglected pairing in the symmetric color sextet channel. In principle,  pairing in the color sextet channel should be included in the  color structure
\cite{Alford:1998mk, Alford:1999pa}, but the condensate in this channel is much smaller than in the $\bar 3$ color channel  \cite{Schafer:1999fe, Shovkovy:1999mr}.

The best studied three-flavor quark phase is  the  so-called color-flavor locked (CFL) phase \cite{Alford:1998mk}, which is believed to be the  energetically favored form of three-flavor quark matter  at asymptotic densities. It is characterized by $\Delta_{1}=\Delta_{2}=\Delta_{3}= \Delta_{\rm CFL}$. The reason of the name ``color-flavor locked"  is that only simultaneous transformations in color and in flavor spaces leave the condensate invariant.  The corresponding  symmetry breaking pattern is the following
\be
  SU(3)_{\it color}\otimes SU(3)_{\it L}\otimes SU(3)_{\it R}\otimes U(1)_B\to SU(3)_{\it color+L+R}
  \otimes Z_2\,,
\label{eq:breakingCFL}
\ee
where $SU(3)_{\it color+L+R}$ is the diagonal global subgroup of the three $SU(3)$ groups. The presence of the $Z_2$ symmetry in the broken phase  means that multiplication of   the quark fields  by  $-1$ leaves the vacuum invariant. 
According to the symmetry breaking pattern, the $17$ generators of   chiral  symmetry,  color symmetry and $U(1)_B$ symmetry are spontaneously broken.  

The 8 broken generators of the color gauge group correspond to the $8$ longitudinal degrees of freedom of the
gluons, thus these gauge bosons acquire a Meissner mass by  the Higgs-Anderson mechanism. According to our Definition \ref{def:superconductor}  the system is a color superconductor. The attribute ``color'' referes to the fact that the broken gauge symmetry is  $SU(3)_{\it color}$. 

The  diquark condensation induces a Majorana-like mass term in the fermionic sector which is not  diagonal in color and flavor indices. Thus, the fermionic excitations consist of  gapped  ``quasiquark" modes, that is modes nondiagonal in color and flavor indices, having mass proportional to $\Delta_{\rm CFL}$. 

The low-energy  spectrum consists of  9 Nambu-Goldstone bosons (NGBs) organized in an  octet, associated with the breaking of the flavor group, and in a singlet, associated with the breaking of the baryonic number, the so called H-phonon. 
 
 For nonvanishing  quark masses the octet of NGBs becomes massive,  but the H-phonon is protected by the  $U(1)_B$ symmetry and remains massless. From our Definition~\ref{def:superfluid} it follows that the CFL phase is a baryonic superfluid.

\section{Crystalline Color Superconductors}
The CFL phase is a particular realization of the standard homogenous BCS pairing, in which fermions of the whole Fermi surface  contribute coherently to the formation of a condensate. In general, in a BCS phase the pairing results in a free energy gain of the order of the homogeneous paring gap, $\Delta_0$. This ideal situation assumes that the Fermi spheres of the interacting fermions  have an equal radius. However, $\beta$-equilibrated  and electrically neutral quark matter might not be such an ideal system. 

For definiteness, let us turn off the strong interactions, meaning that we consider an electro-weakly interacting gas of  $u$, $d$ and  $s$ quarks.
We treat the effective strange quark mass, $M_s$, as a parameter and  consider time scales for which weak equilibrium and electrical neutrality are relevant. The weak processes $u \rightarrow d + \bar e + \nu_e$, $u \rightarrow s + \bar e + \nu_e$ and $u+d \leftrightarrow u + s$ imply that  
\be \mu_u = \mu - \frac{2}{3}\mu_e \qquad \mu_d = \mu +
\frac{1}{3}\mu_e \qquad \mu_s = \mu + \frac{1}{3}\mu_e \,, \label{eq:chemicals-neutral}\ee 
where $\mu= (\mu_u+\mu_d+\mu_s)/3$ is the average chemical potential. The system is (globally)  electrically neutral when the number densities obey the following equation 
\be
\frac{2}3 n_u -\frac{1}3 n_d -\frac{1}3 n_s - n _e=0\,,
\ee
meaning that  the  total electric charge is zero. From the above equations it follows a relation between the electron chemical potential and the effective strange quark mass, see for example \cite{Alford:2002kj},
\be
2\,\delta\mu\equiv\mu_e \simeq \frac{M_s^2}{4 \mu} \,,
\ee
where we have also defined the chemical potential mismatch, $\delta\mu$. 
Certainly, the presence of the strong interaction can change the quark chemical potentials, but the point is that the strong interaction is diagonal in flavor, thus it can only change the intraflavor chemical potential difference between quarks of different colors. Only the locking of the flavor and color  degrees of freedom by a condensate can change significantly the interflavor chemical potential difference. It is possible to show  that the  effect is small and does not significantly change the hierarchy of the chemical potentials~\cite{Anglani:2013gfu}.

\begin{figure}[t]
\begin{center}
\includegraphics[width=12cm]{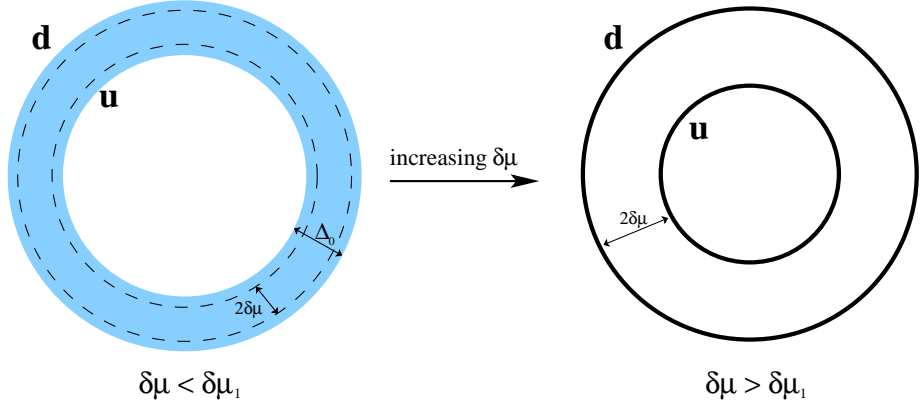}
\end{center}
\caption{\label{Fig:Clogston} Sketch of the Fermi spheres of two populations of fermions, with up ($\bf u$) and down ($\bf d$) spins. Left panel: the dashed black lines correspond to the Fermi spheres of two noninteracting populations for a moderate chemical potential mismatch.  Turning on  the attractive interaction between $\bf u$ and $\bf d$ fermions,   the BCS pairing takes place producing the smearing of the Fermi spheres.  Right panel: for a mismatch above the critical value $\delta\mu_1$ the Fermi spheres of the two populations  are widely separated and the BCS pairing is not energetically favored.}
\end{figure}

If the Fermi surfaces have different radii,  there is a  free energy penalty  proportional to $\delta\mu$ for creating Cooper pairs. For small values of $\delta\mu$, the homogenous BCS phase is still energetically favored. However, it is clear that $\delta\mu$ is producing a  stress on the BCS pairing and this stress cannot be arbitrarily large. In general,   we expect that when $\delta\mu  > c \Delta_0$, where $c$ is some number\footnote{The actual  value of $c$ depends on the detailed form of the interaction; in weak coupling $c=1/\sqrt{2}$ \cite{Chandrasekhar,Clogston}, in strong coupling larger values are allowed, see for example the analysis in \cite{Gubankova:2008ya}.}, the BCS homogenous pairing cannot take place. The naive expectation is that for large  stresses induced by $\delta\mu$ the system makes a phase transition to the normal phase. This situation is pictorially described in Fig.~\ref{Fig:Clogston}. 

The naive picture discussed so far  assumes that only a homogenous condensate can be realized. As frequently happens in physical systems under stress, the transition to inhomogeneous phases might be energetically favored. The first explorations of this direction have been done by  Larkin, Ovchinnikov, Fulde and Ferrell
(LOFF) \cite{LO, FF}.  For a simple system consisting of fermions in two spins states, labeled by ${\it s,t}=1,2$, the FF  condensate  is given by
\begin{equation}
\langle\psi_s(x)\psi_t(x) \rangle\propto  \Delta\, \sigma_{2, {\it st}}\,  e^{2 i \bm q \cdot \bm x}~,
\label{eq:condensate-loff}
\end{equation}
where   $\sigma_2$ is the standard Pauli matrix. The spin structure is characteristic of a spin-0 state, what is new here with respect to BCS pairing is the presence of the plane wave space modulation. In weak coupling this phase is energetically favored with respect to the normal phase in the so-called LOFF window $\delta\mu \in (\delta\mu_1,\delta\mu_2)$, where $\delta\mu_1=\Delta_0/\sqrt{2}$ and $\delta\mu_2\simeq 0.75 \Delta_0$. Let us list  the  properties of  the FF condensate 
\begin{enumerate}[(a)]%
\item \label{prop:def} The condensate is inhomogeneous and is modulated as a plane wave in the $\bm q$ direction.
\item  \label{prop:space} The $SO(3)$ space symmetry is spontaneously broken to an $O(2)$ symmetry, corresponding to rotations around the  $\bm q$ direction.
\item \label{prop:total} The quantity $2 \bm q$ is the total momentum of the pair.
\end{enumerate}
The property \eqref{prop:def} is just a definition of the FF pairing. The property \eqref{prop:space} means that the ground state  is not invariant under space rotations. As a consequence there will be extra low-energy excitations (similar to standard phonons in solids), which are associated with the breaking of the $SO(3)$ space symmetry.  The  property \eqref{prop:total}  (probably the less obvious one\footnote{But the reader can easily prove it by  a Fourier transformation of Eq.~\eqref{eq:condensate-loff}.}),  states that    there is a coherent current of fermionic pairs in one spontaneously chosen direction. However, it does not correspond to a net current of fermions, because there exists a counterpropagating current of unpaired fermions~\cite{FF}. 

\begin{figure}[h!]
\begin{center}
\includegraphics[width=7cm]{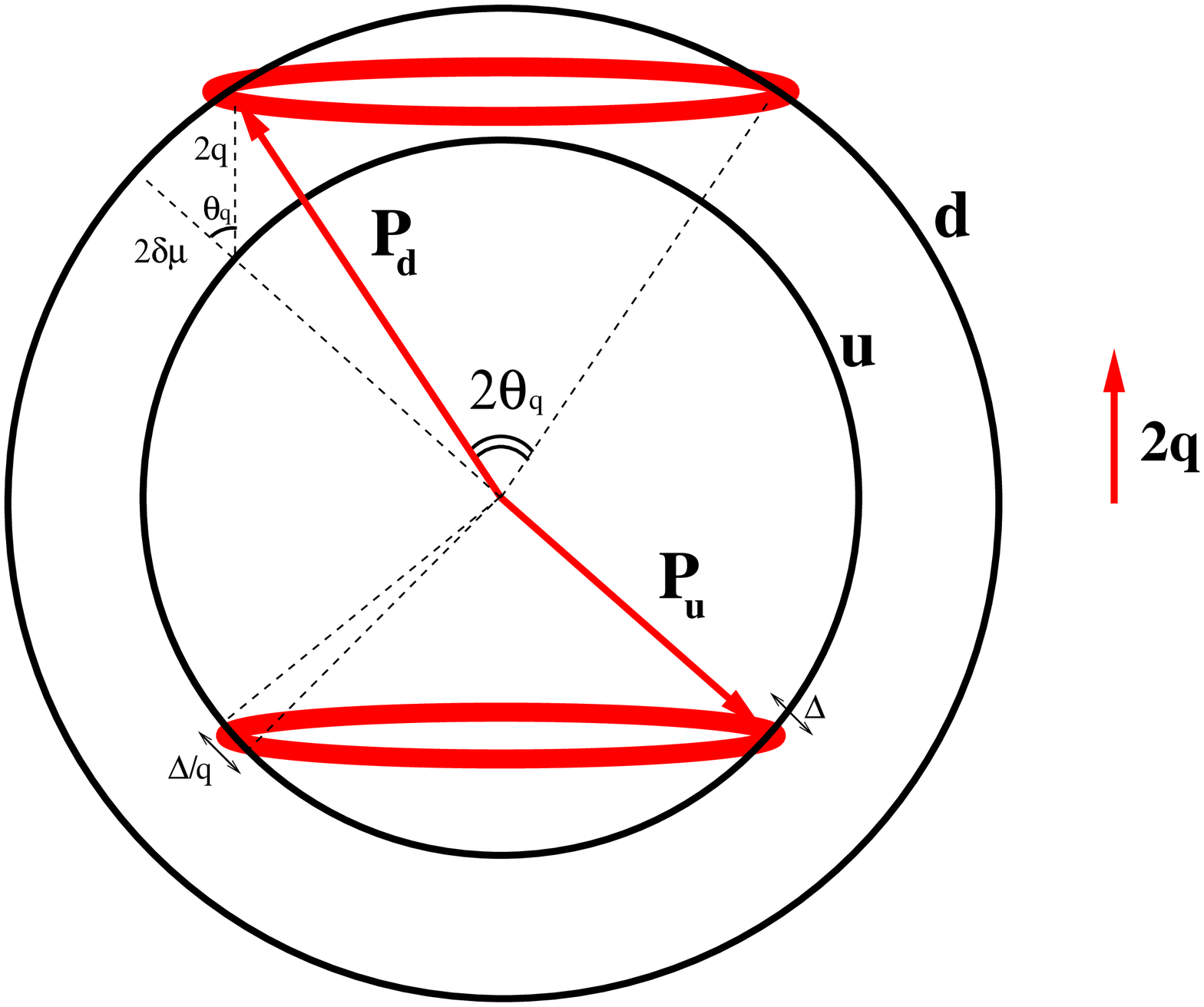}\hspace{1.5cm}
\includegraphics[width=7cm]{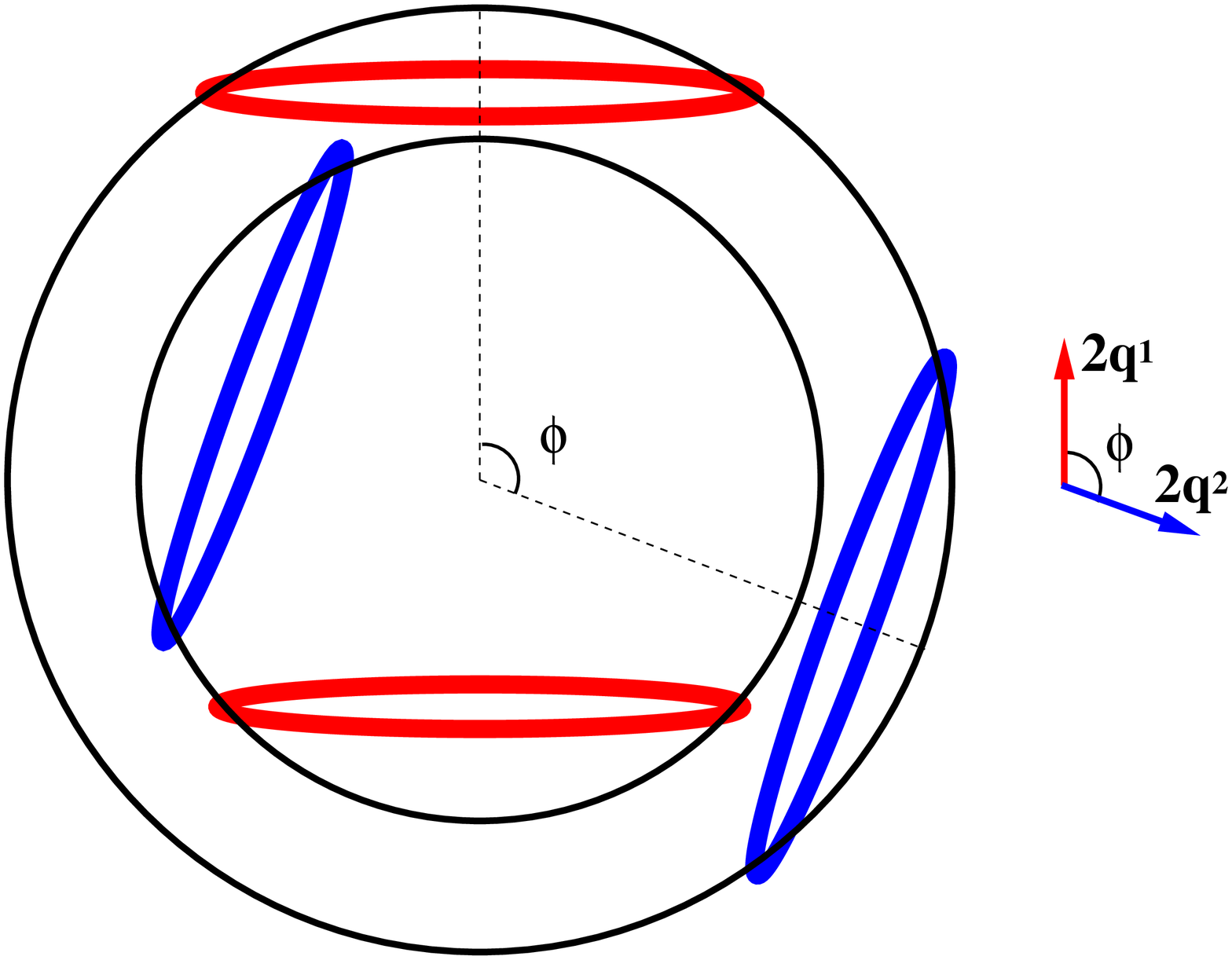}
\end{center}
\caption{\label{Fig:FF}  Pictorial description of the CCSC pairing for two simple structures.  Left panel: In the single plane wave structure pairing takes place in two ribbons  on the top  the Fermi spheres of up and down fermions having opening angle $\arccos( \delta\mu/q) \simeq 67^{\circ}$, thickness $\Delta$ and angular width $\Delta/q$.  Right panel: Structure obtained with two plane waves  with relative angle $\phi$. The size and the opening angle of each ribbon is as in the single plane wave case. More complicated two-flavor CCSC structures can be obtained adding nonoverlapping ribbons on the top of the Fermi spheres. For illustrative purposes, we have greatly exaggerated the splitting between the Fermi surfaces.}
\end{figure}

The Fermi momentum description of the FF pairing is illustrated in the left panel of  Fig.~\ref{Fig:FF}. This figure shows that only a small fraction of the Fermi sphere (corresponding to the two ribbons)   participates in pairing.  These {\it pairing regions} sit on the top of the Fermi spheres, thus there is no free energy price proportional to $\delta\mu$ to be payed. The free energy cost of the FF pairing is instead due to the creation of counterpropagating currents, limiting the extension of the LOFF window. This fact can be qualitatively understood as follows. The value of $q=|\bm q|$ is determined by minimizing the free energy, however it is clear from the geometry of Fig.~\ref{Fig:FF} that $q > \delta\mu$. Thus, increasing $\delta\mu$, $q$ must increase as well, meaning that if the pairing region remains the same  the fermionic current increases. What actually happens, with increasing $\delta\mu$, is that the pairing regions shrink, reducing the  generated current, until they completely disappear at $\delta\mu = \delta\mu_2$. Thus at $\delta\mu_2$ the system makes a smooth (second order) phase transition to the normal phase.

The Fermi sphere description suggests an easy way of generalizing Eq.~\eqref{eq:condensate-loff}, by the addition of a second plane wave, as in the right panel of Fig.~\ref{Fig:FF}. Here $\phi$ indicates the angle between the direction of the two plane waves. The angle $\phi$ can be used as a variational parameter, and it is probably of little surprise the fact that the favored angle corresponds to $\phi = \pi$. Keeping  the discussion qualitative,   in this configuration there are two coherent  counterpropagating currents of paired fermions, therefore no additional current of normal fermions has to be generated. It is also clear that adding more plane waves one can further minimize the free energy, but configurations with overlapping ribbons are disfavored because the associated currents would interfere distroying coherence.

By a further generalization  of the previous reasoning, configurations with several wave vectors corresponding to  nonoverlapping ribbons on the Fermi surfaces should be energetically favored. These wave vectors will thus have an ordered configuration, describing the reciprocal vectors of a crystalline structure. Let us discuss this possibility in quark matter.

Including color and flavor degrees of freedom the generalized color crystalline  condensate can  be obtained by ``joining" Eq.~\eqref{eq:condensate-CFL} with   Eq.~\eqref{eq:condensate-loff}, to give
\begin{equation}
\langle
0|\psi_{iL}^\alpha\psi_{jL}^\beta|0\rangle = - \langle
0|\psi_{iR}^\alpha\psi_{jR}^\beta|0\rangle\propto
  \sum_{I=1}^3 \Delta_I
  \coleps\flaeps \
\sum_{\bm n_{I}^{m}\in \{\bm n_I \}} e^{2i q \bm n_I^m\cdot\bm x} \,,
 \label{eq:condensate-crystal-3}
\end{equation}
where we have suppressed the spinorial indices. 


This condensate describes the crystalline color superconducting (CCSC) phase. Note that the breaking pattern associated with this condensate is the same of the CFL phase \eqref{eq:breakingCFL} with the additional ingredient of the breaking of the rotation symmetries. In most of the crystalline phases the $SO(3)$ space symmetry is  completely broken down to a discrete symmetry characterizing the crystalline pattern.  Any crystalline structure is described by the set $\{\bm n_I \}$, which have an $I$ index because to each of the three interaction channel may correspond a different crystalline arrangement of  vectors.

\begin{figure}[h!]
\includegraphics[width=7cm]{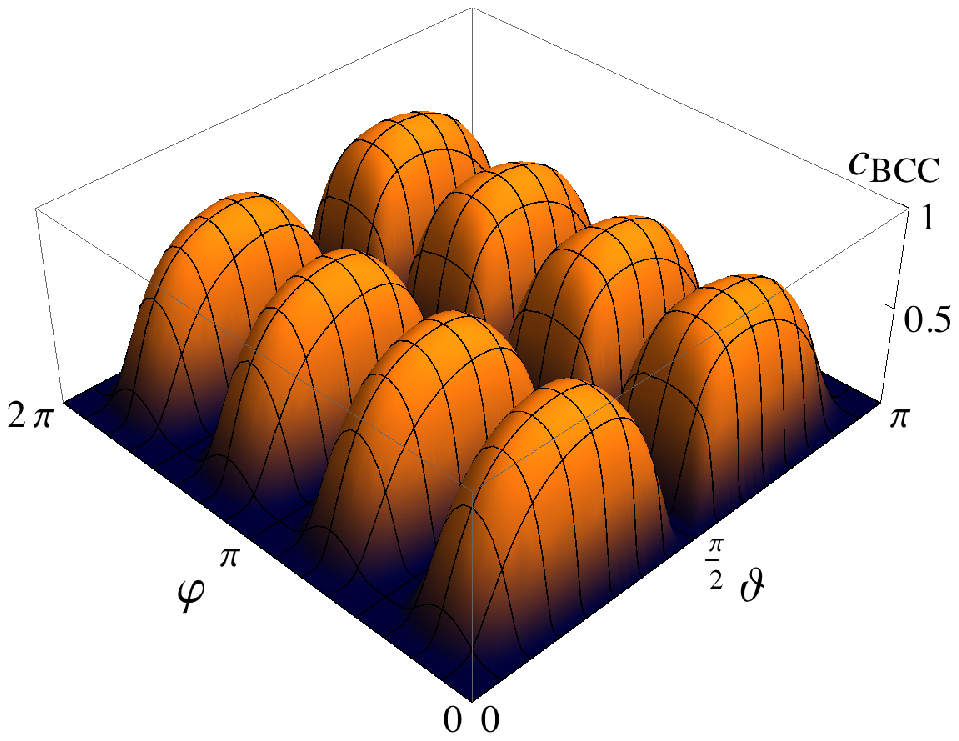}\hspace{1.5cm}
\includegraphics[width=7cm]{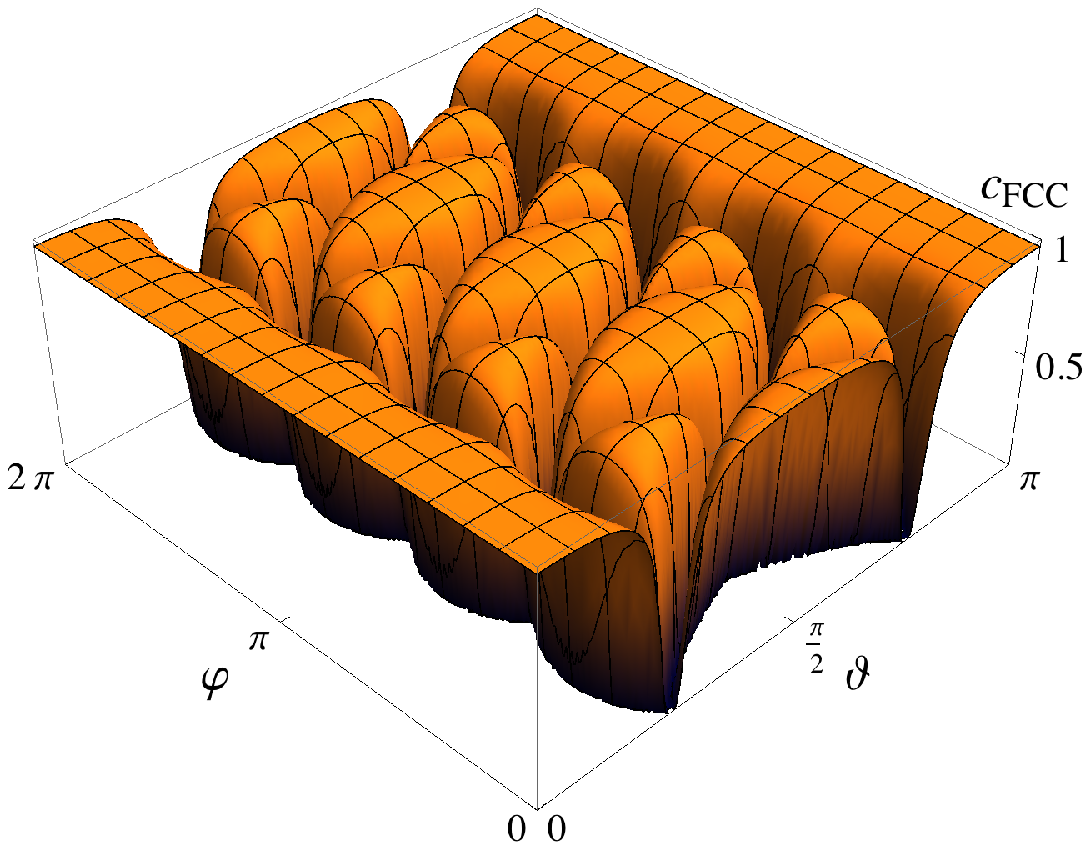}
\caption{\label{fig:speed-of-sound}Velocity of the fermions in the   BCC and FCC two-flavor crystalline phases.}
\end{figure}

The low-energy CCSC spectrum is much richer than the CFL spectrum. We now have not only the H-phonon degree of freedom, but also  $3$ phonon-like modes and $9$ gapless quasiquark modes. Indeed,
an interesting property of the crystalline phase is that for  real-valued periodic condensates  there exists  a fermionic gapless mode  iff  the set $\{\bm n_I \}$ does not contain the null vector,  $\bm n = \bm 0$.  Thus, for these configurations the energy of the fermionic excitations depends linearly on the residual momentum
$\xi$, that is the momentum measured from the Fermi surface. At the lowest order in $\xi/q$  one finds that the quasiquark modes have dispersion law
\be E({\bm v}, \xi) = c({\bm v}) \xi\, , \label{eps1}
\ee 
where $c({\bm v})$ is the direction dependent velocity of the excitations. In  Fig.~\ref{fig:speed-of-sound}  we show  $c({\bm v})$ for two  crystalline structures in two-flavor quark matter with $u$-$d$ paring, that is  with a condensate given in Eq.~\eqref{eq:condensate-crystal-3} with $\Delta_1 = \Delta_2= 0$ and $\Delta_3 \neq 0$ and with the set of vectors $\{\bm n_3\}$ pointing to the vertices of a  BCC structure (left panel) and a FCC structure (right panel). Note that the presence of gapless fermions does not forbid  the existence of superconductivity or superfluidity  \cite{gen66}. Notable condensed matter examples  are type II superconductors, which have gapless fermionic modes for sufficiently large magnetic fields \cite{gen66, Sarma-book}. Note alse that in the Definitions \ref{def:superconductor} and \ref{def:superfluid} there is no mention of the spectrum of fermions.

\begin{figure}[b]
\begin{center}
\includegraphics[width=7.cm,angle=0]{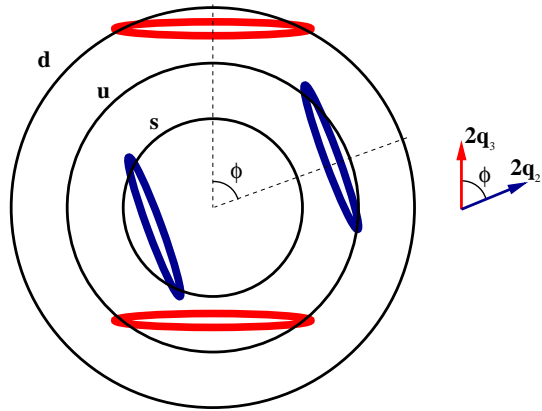}
\end{center}
\caption{(Color online) Sketch showing the hierarchy of the Fermi momenta for three-flavor quark matter and a simple   ``crystal", a particular realization of Eq.~\eqref{eq:condensate-crystal-3}. The  aligned ribbons on the $d$ and $u$ Fermi surfaces indicate those quarks that
contribute to the $\langle ud \rangle$ condensate with gap
parameter $\Delta_3$. The aligned  ribbons on the $u$ and
$s$ Fermi surfaces indicate those quarks that contribute  to the
$\langle us \rangle$ condensate with gap parameter $\Delta_2$. Associated with $\Delta_2$ and $\Delta_3$ are two wave vectors with relative angle  $\phi$. The size and the opening angle of each ribbon is as in Fig.~\ref{Fig:FF}. More complicated three-flavor CCSC structures can be obtained adding nonoverlapping ribbons on the top of the Fermi spheres.  For illustrative purposes we have
greatly exaggerated the splitting between the Fermi surfaces.\label{fig:ribbonsfigure}}
\end{figure}

In three-flavor quark matter there is a richer variety of crystalline structures that can be realized. The number of possibilities is actually too large. One useful simplification is related to the fact that the chemical potential arrangement is as in Fig.~\ref{fig:ribbonsfigure}.  Thus, it is reasonable to expect that $u$-$s$ and $d$-$u$ pairing occurs with almost equal condensates $\Delta_2\simeq \Delta_3 \equiv \Delta$, while the  $d$-$s$ pairing should be suppressed, meaning that  $\Delta_1 \ll \Delta $, and can be neglected. In Fig.~\ref{fig:ribbonsfigure} we illustrate the Fermi spheres of a  simple two plane waves condensate, obtained modulating the $\Delta_2$ condensate by a plane wave with wave vector $\bm q_2$ and the $\Delta_3$ condensate by a plane wave with wave vector $\bm q_3$. Using the argument of the associated currents, it should be easy to understand that  the state with $\phi=0$ is energetically favored. For a proof see \cite{Mannarelli:2006fy,Anglani:2013gfu}. 

Many three-flavor crystalline structures have been considered in~\cite{Rajagopal:2006ig}; among them  the CubeX and the 2Cube45z structures have  the lowest free energy. The CubeX crystal consists of  two sets  $\{\bm n_2\} = \{\bm n_2^1,\bm n_2^2, \bm n_2^3,\bm n_2^4\}$ and $\{\bm n_3\} = \{\bm n_3^1,\bm n_3^2, \bm n_3^3,\bm n_3^4\} $ with
\bea
&\bm n_2^1=\sqrt{\frac{1}3}(1,1,1) =  -\bm n_2^2\,, \qquad & \bm n_2^3=\sqrt{\frac{1}3}(-1,-1,1) =  -\bm n_2^4 \,,\nonumber \\ 
&\bm n_3^1=\sqrt{\frac{1}3}(-1,1,1) =  -\bm n_3^2\,, \qquad & \bm n_3^3=\sqrt{\frac{1}3}(1,-1,1) =  -\bm n_3^4 \,.\nonumber
\eea
Thus, the vectors of each set point to the vertices of a rectangle; the eight vectors together point toward
the vertices of a cube. In the 2Cube45z crystal, $\{\bm n_2\}$ and $\{\bm n_3\}$ each contains
eight wave vectors, pointing to the corners of a cube; the two cubes are rotated 
by $45$ degrees about one of the $4$-fold axes. 

So far, the evaluation of the free energy for these structures has only been done  by means of a Ginzburg-Landau expansion. This method is not quantitatively reliable, however it is useful to have an estimate of the free energy of the various phases. Moreover, at least for some simple structures for which different methods are available, it  underestimates the condensation energy giving a conservative value of the  free energy~\cite{Mannarelli:2006fy}.  The Ginzburg-Landau computations of~\cite{Rajagopal:2006ig} show  that the CCSC phase is favored with respect to the homogeneous CFL and the   unpaired phases in the range
\begin{equation}
2.9\Delta_{\rm CFL} < \frac{M_s^2}{\mu} < 10.4\Delta_{\rm CFL}~. \label{eq:CUBErange}
\end{equation}
Using the self-consistent treatment of~\cite{Ippolito:2007uz} one can transform this range  in 
\be 442\,~{\text MeV} \lesssim \mu\lesssim 515~{\text MeV} ~.\label{cubo}\ee  
This result is certainly model dependent, however it shows that the actual extension of the  LOFF chemical potential window might not be very large.


As a consequence of the spontaneous breaking of the rotation symmetries,  in the low energy spectrum there are phonon-like excitations describing the vibrations  of the crystalline modulation. From the effective Lagrangian describing these modes one can extract the shear modulus of the system \cite{Mannarelli:2007bs}. For the the two favored  structures, 2Cube45z and CubeX, the shear modulus is a $3 \times 3$ nondiagonal matrix in coordinate space with entries proportional to
\begin{equation}
\nu_{\rm CQM} = 2.47\, {\rm MeV}/{\rm fm}^3
\left(\frac{\Delta}{10~{\rm MeV}}\right)^2 \left(\frac{\mu}{400~\rm{MeV}}\right)^2\,.
\label{nunumerical}
\end{equation}
This is a very large  shear modulus, about a  factor of $20-1000$ larger than in 
standard neutron star crust~ \cite{Strohmayer, Mannarelli:2007bs}.

Summarizing, the three-flavor  CCSC phase 
\begin{itemize}
\item Is a color superconductor
\item Is a baryonic superfluid 
\item Has many low energy degrees of freedom:  the H-phonon, the  phonon-like modes (bosonic), the quark quasiparticles (fermionic)
\item Is characterized by an extremely rigid space modulation of the condensate
\item Is a supersolid. 
\end{itemize}

The last property follows from the Definition \ref{def:supersolids}. In the CCSC phase there is indeed the simultaneous breaking of the gauge color symmetry, of the global $U(1)_B$ symmetry and of the rotational symmetries. These symmetry breakings happen simultaneously because of the condensate in Eq.~\eqref{eq:condensate-crystal-3}.

\section{Astrophysics}
Given the exceptional properties of the CCSC phase, one would naively expect being easy to verify whether or not this phase is realized in nature. However, testing the existence  of color superconductors is extremely nontrivial. The region of  low-temperature and high-density is not easily reachable in terrestrial  high-energy experiments \cite{Klahn:2012uq}. Thus, to date the only few indirect information which we could put in relation with the color superconducting phase  come from astronomical observations of compact stellar objects (CSOs). 

The reason we expect color superconductors in CSOs is the following. In CSOs   matter is compressed at densities about a factor $3$-$5$ larger than in a heavy nucleus. A simple geometrical reasoning suggests that   baryons are likely to lose their identity and dissolve into deconfined quarks~\cite{Weber-book}. In this case  compact (hybrid) stars featuring quark cores  would exist. A different possibility is that 
$uds$  quark  matter is the ground state of the hadrons~\cite{Witten:1984rs}. In this case, the so-called strange stars~\cite{Alcock:1986hz}   completely composed of deconfined matter should exist.

In CSOs, quark matter is long-lived, charge neutral and in $\beta$-equilibrium, which we have seen are good conditions for the realization of the CCSC phase. Moreover, the critical temperature (in weak coupling) is given  by $T_{c} \simeq 0.57 \Delta$ and reasonable values  of the CCSC gap parameter range in between $5$ MeV and $25$ MeV~\cite{Mannarelli:2007bs}. Since for the greatest part of the CSO lifetime the temperature is much lower than this critical temperature the CCSC state should be thermodynamically favored. 

Recent astronomical observation of very massive CSOs \cite{Demorest:2010bx} seem to disfavor the possibility that a color superconducting phase is realized~\cite{Logoteta:2012ms}, but the results depend on the poorly known equation of state  of matter at high density, and the possibility that hybrid stars of about  $2 M_\odot$ have a CCSC core  \cite{Ippolito:2007hn,Anglani:2013gfu} cannot be excluded.  Basically the available mass-radius observations  do not allow us to infer in a unique way  the internal structures of CSOs because  hybrid stars featuring quark matter could masquerade as standard neutron stars~\cite{Alford:2004pf}. 

Any other astronomical observation, related to cooling,  radial and nonradial oscillations, glitches, strong magnetic fields etc. is based on the poorly known internal structure of the star and on its evolution after its birth. Therefore,  it is very hard to extract useful and model independent  information from these observations. Given the large uncertainty in the estimate of the parameters characterizing the CCSC phase, to date astronomical  observations could only been used to restrict the parameter space of the model.  Nevertheless,  the investigation of these astrophysical signatures is the only means we have to connect theoretical models with astronomical observations.

One of the most promising direction is the study of the astrophysical properties related to the extraordinary rigidity  of the CCSC phase. This is indeed a peculiar property of the CCSC phase; no other phase is known to have a comparable shear modulus.  Suppose that the CCSC phase is realized in the core of a CSO. Then,  the rigidity  of the CCSC phase may allow the presence of   large deformation of the core.   If this deformation has a  nonzero quadrupole moment  with respect to the rotation axis  and if this axis-asymmetric mass distribution is not compensated by the overlying nuclear envelope,  a spinning CSO  would efficiently emit  gravitational waves  \cite{Lin:2007rz, Haskell:2007zz,Anglani:2013gfu}. A second possibility is that  the emission of gravitational waves happens by excitation of the torsional oscillations~\cite{Lin:2013nza}.  These are  particular toroidal oscillations, see for example \cite{1988ApJ...325..725M}, which when   triggered (say by a stellar glitch), produce an emission of  gravitational waves with an emitted power roughly proportional to $\nu^4$ with frequencies in the kHz range. In principle these gravitational waves  could be detected by the second-generation gravitational-wave detectors. A positive result would be, for many reasons, a great achievement and a clear indication of the existence of the CCSC phase. A negative result would be useful for further reducing the parameter space of the CCSC phase.

\ack
I acknowledge discussion with G.~Pupillo, who triggered my interest in supersolids and  the realization of the supersolid phase in compact stars.

\section*{References}
\bibliographystyle{iopart-num}
\providecommand{\newblock}{}


\end{document}